\documentclass[11pt]{article}
\usepackage{mathptmx}
\usepackage{amssymb}
\usepackage[dvips]{color}
\usepackage[hyphens]{url}
\usepackage{graphicx}
\usepackage[german,english]{babel}
\usepackage{soul}

\long\def\symbolfootnote[#1]#2{\begingroup%
\def\thefootnote{\fnsymbol{footnote}}\footnote[#1]{#2}\endgroup} 

\title{A Critical Look at the Standard Cosmological Picture}
\author{Daryl Janzen\footnote{\textrm{email: daryl.janzen@usask.ca}}}
\date{}
\begin{document}
\maketitle
\begin{abstract}
The discovery that the Universe is accelerating in its expansion has brought the basic concept of cosmic expansion into question. An analysis of the evolution of this concept suggests that the paradigm that was finally settled into prior to that discovery was not the best option, as the observed acceleration lends empirical support to an alternative which could incidentally explain expansion in general. I suggest, then, that incomplete reasoning regarding the nature of cosmic time in the derivation of the standard model is the reason why the theory cannot coincide with this alternative concept. Therefore, through an investigation of the theoretical and empirical facts surrounding the nature of cosmic time, I argue that an enduring three-dimensional cosmic present must necessarily be assumed in relativistic cosmology---and in a stricter sense than it has been. Finally, I point to a related result which could offer a better explanation of the empirically constrained expansion rate.
\end{abstract}
\section{Introduction \label{sec_I}}
Many of our basic conceptions about the nature of physical reality inevitably turn out to have been false, as novel empirical evidence is obtained, or paradoxical implications stemming from those concepts are eventually realised. This was expressed well by Einstein, who wrote \cite{Einstein1916}
\begin{quotation}
What is essential, which is based solely on accidents of development?\ldots Concepts that have proven useful in the order of things, easily attain such an authority over us that we forget their Earthly origins and accept them as unalterable facts.\ldots 
The path of scientific advance is often made impassable for a long time through such errors. It is therefore by no means an idle trifling, if we become practiced in analysing the long-familiar concepts, and show upon which circumstances their justification and applicability depend, as they have grown up, individually, from the facts of experience. 

\end{quotation}
Or, as he put it some years later \cite{Einstein1931},
\begin{quotation}
The belief in an external world independent of the percipient subject is the foundation of all science. But since our sense-perceptions inform us only indirectly of this external world, or Physical Reality, it is only by speculation that it can become comprehensible to us. From this it follows that our conceptions of Physical Reality can never be definitive; we must always be ready to alter them, to alter, that is, the axiomatic basis of physics, in order to take account of the facts of perception with the greatest possible logical completeness.
\end{quotation}
And so it is in the same spirit, that I shall argue against a number of concepts in the standard cosmological picture that have changed very little in the past century, by making note of original justifications upon which they were based, and weighing those against empirical data and theoretical developments that have been realised through the intervening years.

The essay will concentrate initially on the nature of cosmic expansion, which lacks an explanation in the standard cosmological model. Through a discussion of the early developments in cosmology, a familiarity with the pioneering conception of expansion, as being always driven by a cosmological constant $\Lambda$, will be developed, upon which basis it will be argued that the standard model---which cannot reconcile with this view---affords only a very limited description. Then, the nature of time in relativistic cosmology will be addressed, particularly with regard to the formulation of `Weyl's postulate' of a cosmic rest-frame. The aim will therefore be towards a better explanation of cosmic expansion in general, along with the present acceleration that has recently become evident, by reconceiving the description of time in standard cosmology, as an approach to resolving this significant shortcoming of the big bang Friedman-Lema\^{i}tre-Robertson-Walker (FLRW) models, and particularly the flat $\Lambda$CDM model that describes the data so well.

\section{On Cosmic Expansion \label{sec_OCE}}
The expansion of our Universe was first evidenced by redshift measurements of spiral nebulae, after the task of measuring their radial velocities was initiated in 1912 by Slipher; and shortly thereafter, 
de~Sitter attempted the first relativistic interpretation of the observed shifts, noting that `the frequency of light-vibrations diminishes with increasing distance from the origin of co-ordinates' due to the coefficient of the time-coordinate in his solution \cite{deSitter1917}. But the concept of an expanding Universe, filled with island galaxies that would all appear to be receding from any given location at rates increasing with distance, was yet to fully form.

For one thing, when de~Sitter published his paper, he was able to quote only three reliable radial velocity measurements, which gave merely $2:1$ odds in favour of his prediction. However, in 1923 Eddington produced an updated analysis of de~Sitter space, and showed that the redshift de~Sitter had predicted as a phenomenon of his statical geometry was in fact due to a cosmical repulsion brought in by the $\Lambda$-term, which would cause inertial particles to all recede exponentially from any one \cite{Eddington1923}. He used this result to support an argument for a truly expanding Universe, which would expand everywhere and at all times due to $\Lambda$. This, he supported with an updated list of redshifts from Slipher, which now gave $36:5$ odds in favour of the expansion scenario.

That same year, Weyl published a third appendix to his \textit{Raum, Zeit, Materie}, and an accompanying paper \cite{Weyl1923}, where he calculated the redshift for the `de~Sitter cosmology', 
\begin{equation}
\mathrm{d}s^2=-\mathrm{d}t^2+e^{2\sqrt{\frac{\Lambda}{3}}t}(\mathrm{d}x^2+\mathrm{d}y^2+\mathrm{d}z^2),\label{dS_LR}
\end{equation}
the explicit form of which would only be found later, independently by Lema\^{i}tre \cite{Lemaitre1925b} and Robertson \cite{Robertson1928}. Weyl was as interested in the potential relevance of de~Sitter's solution for an expanding cosmology as Eddington \cite{Weyl1923}, and had indeed been confused when he received a postcard from Einstein later that year (Einstein Archives: [24-81.00]), stating,
\begin{quotation}
With reference to the cosmological problem, I am not of your opinion. Following de~Sitter, we know that two sufficiently separate material points are accelerated from one another. If there is no quasi-static world, then away with the cosmological term.
\end{quotation}
Eight days after this was posted, Einstein's famous second note \cite{Einstein1923} on Friedman's paper, which he now referred to as `correct and clarifying', arrived at \textit{Zeitschrift f\,\"{u}r Physik}. Einstein evidently had in mind that the cosmic expansion can be described with $\Lambda$ set to zero in Friedman's solution, and he might have thought Weyl would notice \cite{Einstein1923} and make the connection---but the latter evidently did not, as he wrote a dialogue the following year \cite{Weyl1924} in which the proponent of orthodox relativity\footnote{The dialogue is set between Saints Peter and Paul, with the latter presenting Weyl's `apostatical' and `heretical' views against the `Relativity Church'. The following statement, which seems to be loosely quoted from the postcard sent by Einstein, was made by Peter.} eventually states, `If the cosmological term fails to help with leading through to Mach's principle, then I consider it to be generally useless, and am for the return to the elementary cosmology'---that being a particular foliation of Minkowski space, which, of the three cosmological models known to Weyl, was the only one with vanishing $\Lambda$.

\begin{figure}[t!]
\centerline{\includegraphics[height=8.7cm]{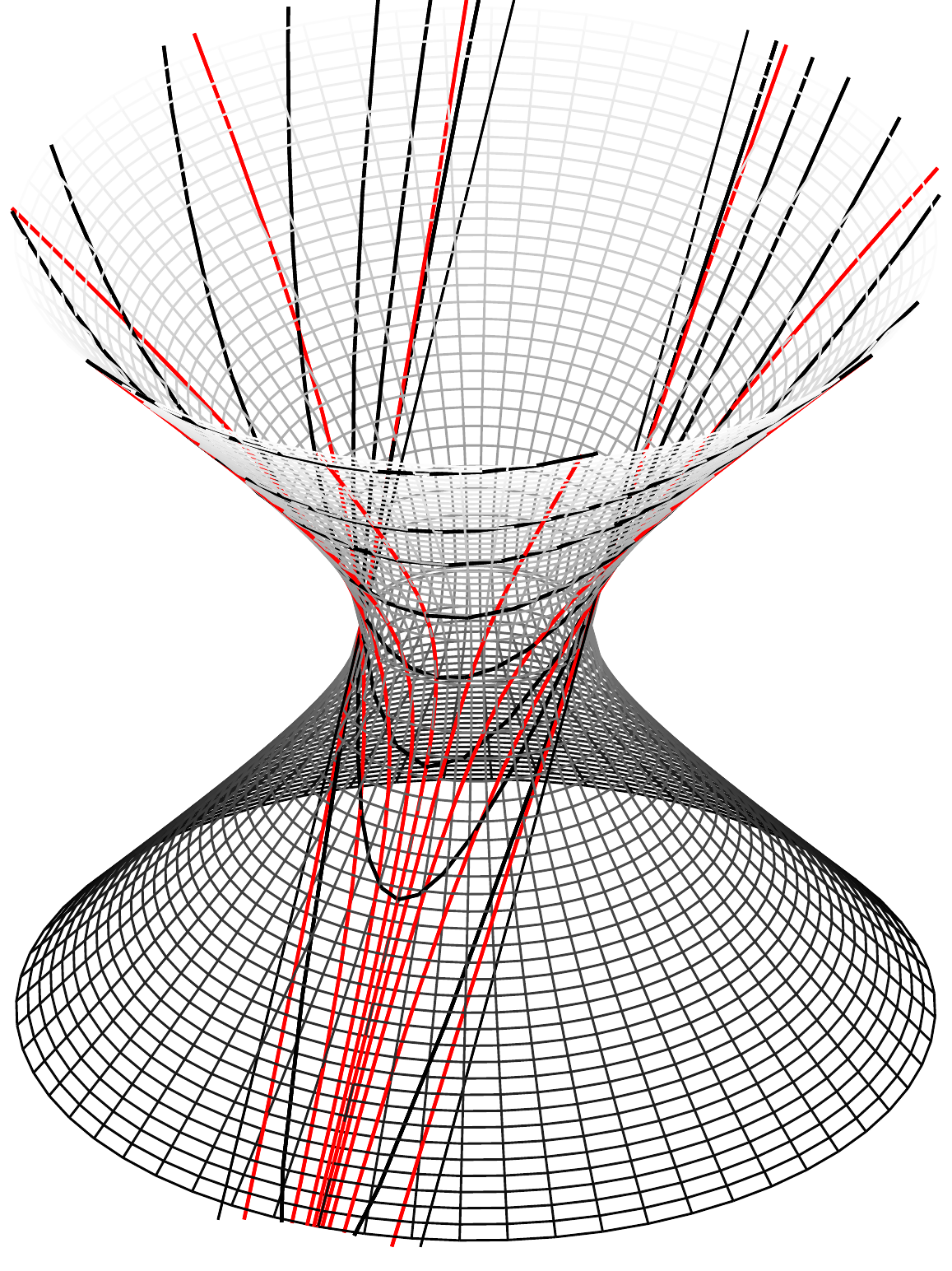}\includegraphics[height=8.7cm]{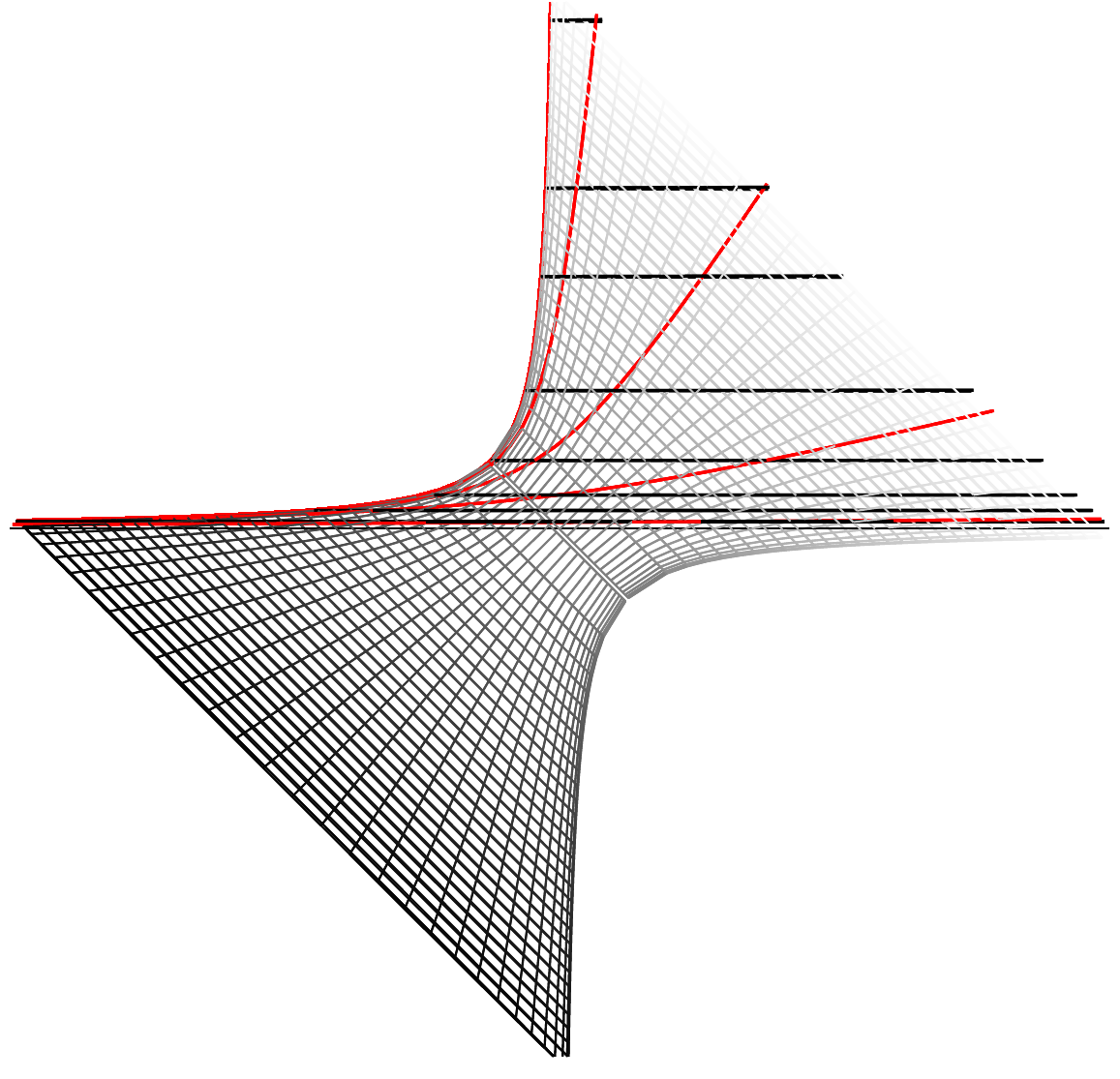}}
\caption{Slices of constant time in the Lema\^{i}tre-Robertson coordination of de~Sitter space (black lines), along with comoving world lines (red lines), drawn on a two-dimensional slice of de~Sitter space in three-dimensional Minkowski space.}\label{fig: dS_LR}
\end{figure}

At this point in the dialogue, the protagonist Paulus perseveres, citing the evidence for an expanding Universe, and therefore the de~Sitter cosmology as the most likely of the three known alternatives. Weyl's excitement over its description is evident in Paulus' final statement: `If I think about how, on the de~Sitter hyperboloid the world lines of a star system with a common asymptote rise up from the infinite past [see Fig.~\ref{fig: dS_LR}], then I would like to say: the World is born from the eternal repose of `Father {\AE}ther'; but once disturbed by the `Spirit of Unrest' (\textit{H\"{o}lderlin}), which is at home in the Agent of Matter, `in the breast of the Earth and Man', it will never come again to rest.' Indeed, as Eq.~(\ref{dS_LR}) indicates, and as illustrated in Fig.~\ref{fig: dS_LR}, the universe emerges from a single point at $t=-\infty$, even though slices of constant cosmic time are infinitely extended thereafter---and comoving geodesics \textit{naturally} disperse throughout the course of cosmic time.

Thus, we have a sense of the concept of cosmic expansion that was common amongst the main thinkers in cosmology in the 1920s, who were considering the possibility of expansion driven by the cosmical repulsion in de~Sitter space. Indeed, Hubble was aware of this concept, as he wrote of the `de~Sitter effect' when he published his confirmation of cosmic expansion in 1929 \cite{Hubble1929}; and de~Sitter himself, in 1930, wrote of $\Lambda$ as `a measure of the inherent expanding force of the universe' \cite{deSitter1930}. Thus, along with the evidence that our Universe actually \textit{does} expand, one had in-hand the description of a well-defined force to \textit{always} drive that expansion.

It was therefore a huge blow to Eddington, e.g., when in 1932 Einstein and de~Sitter \cite{Einstein1932} finally rejected that interpretation of cosmic expansion, in favour of a model that could afford no prior explanation for \textit{why} the Universe \textit{should} expand. As he put it \cite{Eddington1933},
\begin{quote}
the theory recently suggested by Einstein and de~Sitter, that in the beginning all the matter created was projected with a radial motion so as to disperse even faster than the present rate of dispersal of galaxies,\symbolfootnote[1]{They do not state this in words, but it is the meaning of their mathematical formulae. [Eddington's footnote.]} leaves me cold. One cannot deny the possibility, but it is difficult to see what mental satisfaction such a theory is supposed to afford.
\end{quote}

To see why the big bang FLRW models with matter provide no explanation of expansion, for the reason stated by Eddington, we need only look at Friedman's equation,
\begin{equation}
\frac{\ddot{a}}{a}=\frac{\Lambda}{3}-\frac{\kappa}{2}\left(p+\frac{\rho}{3}\right),\label{ddot_a}
\end{equation} 
which describes the dependence of the scale-factor, $a$, on $\Lambda$ and the density, $\rho$, and pressure, $p$, of matter. Since $p+\rho/3$ goes like $1/a^4$ for radiation or $1/a^3$ for non-relativistic matter, the \textit{decelerative} force due to finite matter-densities blows up exponentially as $a\rightarrow0$, while the \textit{accelerative} force due to $\Lambda$ vanishes; so the `inherent expanding force of the universe' only contributes to the expansion of space later on, when the relative contributions of matter and radiation have sufficiently weakened. Therefore, aside from Weyl's vacuous de~Sitter cosmology, with its big bang singularity at $t=-\infty$, the big bang FLRW models can never \textit{explain} the cosmic expansion they describe, which must be caused by the big bang singularity itself---i.e., where the theory blows up. 

But since the cosmic microwave background radiation (CMBR) indicates that the Universe \textit{did} begin in a hot dense state at a finite time in the past, the model Eddington had favoured instead (in which an unstable Einstein universe that existed since eternity would inevitably begin expanding purely due to $\Lambda$ \cite{Eddington1930}) also can't be accepted.

The principal source of standard cosmology's great \textit{explanatory deficit} is the fact that although the non-vacuous big bang FLRW models do \textit{describe} expanding universes---and in particular the flat $\Lambda$CDM model describes the observed expansion of our Universe very well \cite{SNe Ia}---they afford no reason at all for \textit{why} those universes \textit{should} expand, since that could only be due to the initial singularity; i.e., as we follow the models back in time, looking for a possible cause of expansion, we eventually reach a point where the theory becomes undefined, and call that the cause of it all. In contrast, I've discussed two FLRW models, neither of which is empirically supported, which would otherwise better \textit{explain} the expansion they describe, as the result of a force that is well-defined in theory.

The basic cause and nature of cosmic expansion, along with its recently-observed acceleration, are significant problems of the standard model; so, condisering the evidence that the acceleration is best described by pure $\Lambda$ \cite{SNe Ia}, there is strong motivation to search for an alternative big bang model that would respect the pioneering concept of expansion, as a direct consequence of the `de~Sitter effect' in the modified Einstein field equations. It is therefore worth investigating the axiomatic basis of the Robertson-Walker (RW) line-element. As I will eventually argue that the problem lies in the basic assumptions pertaining to the description of cosmic time, I'll begin by discussing some issues related to the problem of accounting for a cosmic present.

\section{The Cosmic Present \label{sec_TCP}}
The problem of recognising a cosmic present is that, according to relativity theory, it should not be possible to assign one time-coordinate to the four-dimensional continuum of events that could be used to describe objective simultaneity, since two events that are described as simultaneous in one frame of reference will not be described as such by an observer in relative motion. However, as noted by Bondi \cite{Bondi1960},
\begin{quotation}
The Newtonian concept of the uniform omnipresent even-flowing time was shown by special relativity to be devoid of physical meaning, but in 1923 H.\ Weyl suggested that the observed motions of the nebulae showed a regularity which could be interpreted as implying a certain geometrical property of the substratum \ldots. This in turn implies that it is possible to introduce an omnipresent \textit{cosmic time} which has the property of measuring \textit{proper time} for every observer moving with the substratum. In other words, whereas special relativity shows that a set of arbitrarily moving observers could not find a common `time', the substratum observers move in such a specialized way that such a public or cosmic time exists.

Although the existence of such a time concept seems in some ways to be opposed to the generality, which forms the very basis of the general theory of relativity, the development of relativistic cosmology is impossible without such an assumption.
\end{quotation}
In fact, as Einstein himself noted in 1917 \cite{Einstein1917}, 
\begin{quotation}
The most important fact that we draw from experience as to the distribution of matter is that the relative velocities of the stars are very small as compared with the velocity of light. So I think that for the present we may base our reasoning upon the following approximative assumption. There is a system of reference relatively to which matter may be looked upon as being permanently at rest.
\end{quotation}
Thus, the assumption of a cosmic rest-frame---and a corresponding cosmic time---was justified in the derivation of Einstein's `cylindrical' model.

While Einstein originally proposed this as an `approximative assumption' that the empirical evidence seemed to support, the fact that he did restore absolute time when it came to the problem of describing the Universe on the largest scale was not lost on his peers. De~Sitter was immediately critical of the absolute time variable in Einstein's model, noting that `Such a fundamental difference between the time and the space-coordinates seems to be somewhat contradictory to the complete symmetry of the field-equaitons and the equations of motion' \cite{deSitter1917a}. And a few years later, Eddington wrote that an objection to Einstein's theory may be urged, since \cite{Eddington1920} `absolute space and time are restored for phenomena on a cosmical scale\ldots Just as each limited observer has his own particular separation of space and time, so a being coexistive with the world might well have a special separation of space and time natural to him. It is the time for this being that is here dignified by the title ``absolute.{''}' Therefore, he concluded, `Some may be inclined to challenge the right of the Einstein theory\ldots to be called a relativity theory. Perhaps it has not all the characteristics which have at one time or another been associated with that name\ldots'

Indeed, although the assumption of an absolute time in relativistic cosmology is definitely not in the spirit of relativity, the theory isn't fundamentally incompatible with such a definition. Furthermore, it is significant that despite such early criticisms, Einstein never wavered in assuming an absolute time when he came to consider the cosmological problem \cite{Einstein1931b,Einstein1932,Einstein1945}, i.e. as he always favoured the Friedman solutions (with $\Lambda=0$), which begin by postulating the same.

So, we have two opposing descriptions of relativistic time---both of which are principally due to Einstein himself!---and what I'll now argue is that developments both in cosmology and in our understanding of relativity theory which have taken place in the past century demand the latter---that there is one absolute cosmic time relative to which every observer's proper time will measure, as space-time will be perceived differently due to their absolute motion through the cosmic present that must be uniquely and objectively defined---rather than the former implication of Einstein's 1905 theory of relativity \cite{Einstein1905}. 

In the case of special relativity, a description in which space-time emerges as a clearly defined absolute cosmic present endures, can be realised by considering four-dimensional Minkowski space, as a background structure, and a three-dimensional universe that actually flows equably though it---with the past space-time continuum emerging as a purely ideal set of previous occurrences in the universe. Then, if we begin in the cosmic rest-frame, in which fundamental observers' world lines will be traced out orthogonal to the cosmic hyperplane, photons can be described as particles that move through that surface at the same rate as cosmic time, thus tracing out invariant null-lines in space-time. In this way, the evolution of separate bodies, all existing in one three-dimensional space, forms a graduating four-dimensional map. 

The causal and inertial structures of special relativity are thus reconciled by describing the world lines of all observers in uniform motion through the cosmic present as their proper time axes, and rotating their proper spatial axes accordingly, so that light will be described as moving at the same rate in either direction of proper `space'. And then, so that the speed of photons along invariant null-lines will actually be the same magnitude in all inertial frames, both the proper space and time axes in these local frames must also be scaled hyperbolically relative to each other.

This description of the emergence of space-time in a special relativistic universe can be illustrated in the following way. Consider a barograph, consisting of a pen, attached to a barometer, and a sheet of paper that scrolls under the pen by clockwork. The apparatus may be oriented so that the paper scrolls downwards, with changes in barometric pressure causing the pen to move purely horizontally. We restrict the speed of the pen's horizontal motion only so that it must always be less than the rate at which the paper scrolls underneath it. The trace of the barometric pressure therefore represents the world line of an arbitrarily moving observer in special relativistic space-time, with instantaneous velocity described in this frame by the ratio of its speed through the horizontal cosmic present and the graph paper's vertical speed, with `speed' measured in either case relative to the ticking of the clockwork mechanism, which therefore cancels in the ratio.

Now, in order to illustrate the relativity of simultaneity, we detach the pen (call it $\mathcal{A}$) from the barometer so that it remains at rest absolutely, and add another pen, $\mathcal{B}$, to the apparatus, at the exact same height, which moves horizontally at a constant rate that's less than the constant rate that the paper scrolls along; therefore, with \textit{absolute velocity} less than the absolute speed limit. Furthermore, we make $\mathcal{A}$ and $\mathcal{B}$ `observers', by enabling them to send and receive signals that transmit horizontally at the same rate (in clockwork time) as absolute time rolls on (in clockwork time), thus tracing out lines on the graph paper with unit speed.

As this system evolves, the two `timelike observers' can send these `photons' back and forth while a special relativistic space-time diagram is traced out. If we'd rather plot the map of events in coordinates that give the relevant description from $\mathcal{B}$'s perspective, we use the Lorentz transformation equations corresponding to the description of the map as Minkowski space-time: a spacelike line is drawn, tilted from the horizontal towards $\mathcal{B}$'s world line by the appropriate angle, and the events along that surface are described as synchronous in that frame, even though they take place sequentially in real time. In particular, at the evolving present, $\mathcal{B}$'s proper spatial axis extends, in one direction, onto the empty sheet of graph paper in which events have not yet occurred, and, in the other direction, into the past space-time continuum of events that have already been traced onto the paper---while the real present hyperplane, where truly simultaneous events are occurring, is tilted with respect to that axis of relative synchronicity.

The main difference between this interpretation of special relativity and Einstein's original one, is that `simultaneity' and `synchronicity' have objectively different meanings for us, which coincide only in the absolute rest frame---whereas Einstein established an `operational' concept of simultaneity, so that it would be synonimous with synchronicity, in section 1, part 1 of his first relativity paper \cite{Einstein1905}. Einstein's definition of simultaneity is a basic assumption that's really no less arbitrary than Newton's definitions of absolute space, time, and motion; and, as I'll argue, the evidence from cosmology now stands against Einstein's wrong assumption, as it is really more in line with Newton's.

\begin{figure}[t!]
\centerline{\includegraphics[height=4.25cm]{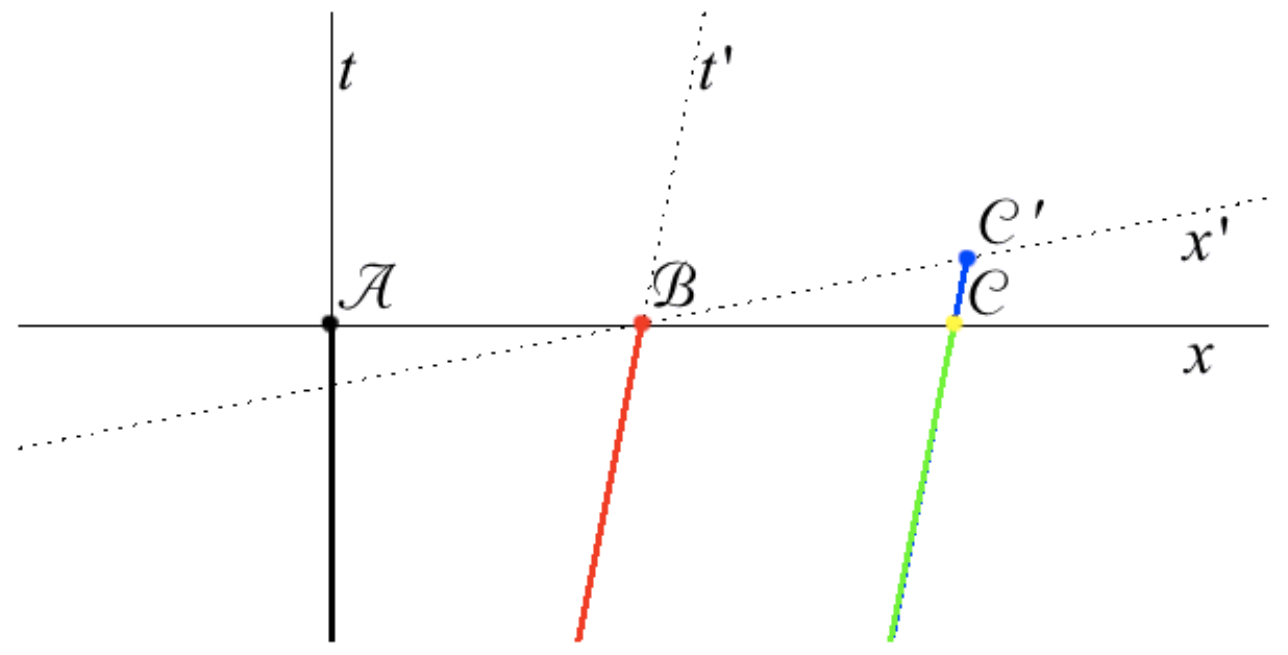}\includegraphics[height=4.25cm]{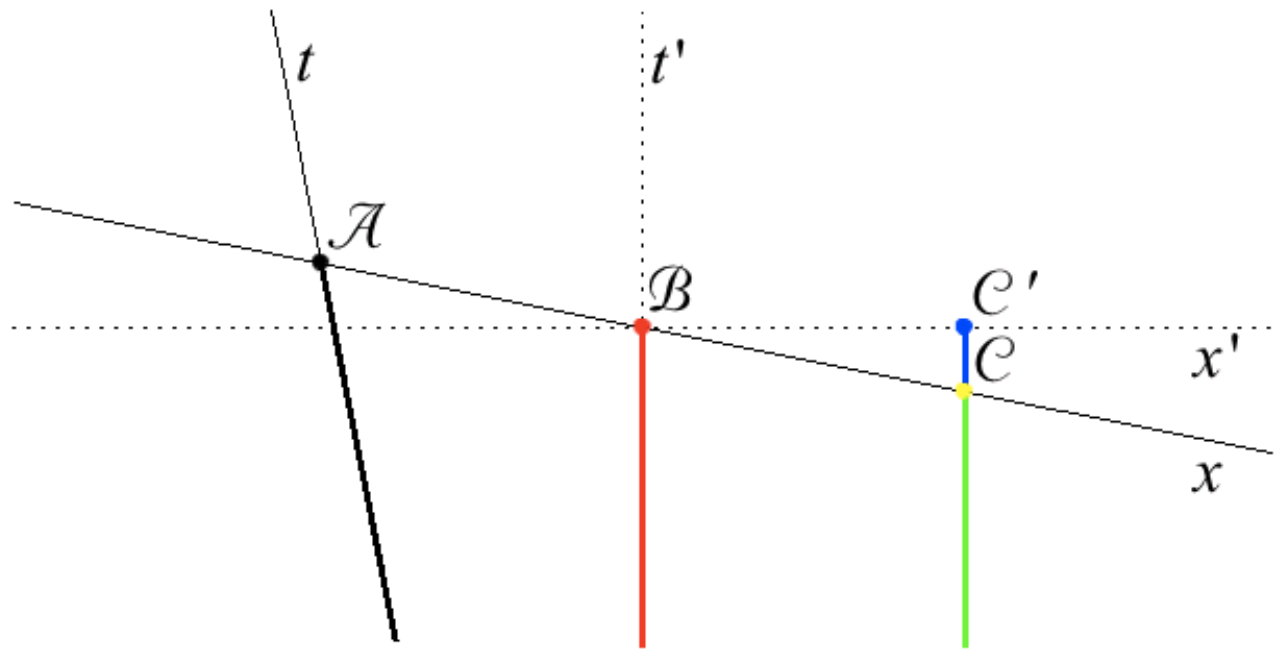}}
\caption{Snapshots, in two proper reference frames, of an emergent space-time. Although the proper times of $\mathcal{C'}$ and $\mathcal{B}$ appear to coincide, $\mathcal{C'}$ is disconnected from the causally coherent set, $\{\mathcal{A},\mathcal{B},\mathcal{C}\}$.}\label{fig: Elementary cosmology}
\end{figure}

The distinction between simultaneity and synchronicity in this different interpretation of relativity, can be understood more clearly through our barograph example, by adding two more `observers', $\mathcal{C}$ and $\mathcal{C}'$, which remain at rest relative to $\mathcal{B}$, with $\mathcal{C}$ positioned along the same hyperplane as $\mathcal{A}$ and $\mathcal{B}$, and $\mathcal{C}'$ positioned precisely at the intersection of $\mathcal{C}$'s world line (so that the world lines of $\mathcal{C}$ and $\mathcal{C}'$ exactly coincide, as they are traced out on the space-time graph) and $\mathcal{B}$'s proper spatial axis (therefore, on a different hyperplane than $\mathcal{A}$, $\mathcal{B}$, and $\mathcal{C}$); thus, $\mathcal{C}'$ shall not be causally connected to $\mathcal{A}$, $\mathcal{B}$, and $\mathcal{C}$, since \textit{by definition} information can only transmit along the cosmic hyperplane; see Fig.~\ref{fig: Elementary cosmology}. 

The significant point that is clearly illustrated through the addition of $\mathcal{C}$ and $\mathcal{C}'$, is that although in the proper coordinate system of $\mathcal{B}$ (or $\mathcal{C}$ or $\mathcal{C'}$), $\mathcal{C'}$ appears to exist synchronously and at rest relative to $\mathcal{B}$, $\mathcal{C}$---which in contrast appears to exist in $\mathcal{B}$'s (spacelike separated) past or future (depending on the direction of absolute motion; in Fig.~\ref{fig: Elementary cosmology} $\mathcal{C}$ appears to exist in $\mathcal{B}$'s relative past)---is really the causally connected neighbour that remains relatively at rest, with which it should be able to synchronise its clock in the usual way; i.e., the synchronisation of $\mathcal{B}$'s and $\mathcal{C}$'s clocks will be \textit{wrong} because \textit{simultaneous noumena will not be perceived as synchronous phenomena in any but the cosmic rest-frame}.

According to this description, we should have to relinquish the concept that there can be no priviliged observers, as well as Einstein's light-postulate in its original form. With regard to the latter, consider that photons will still be described as travelling at a constant speed in all directions of all reference frames, due to the invariance of null-lines. But this won't actually be true, since an observer moving through the universe will keep pace better with a photon in their direction of motion, and will remain closer to that photon at all later times, on the cosmic hyperplane. Therefore, although light actually won't recede as quickly through the universe in the direction of absolute motion, it can always be described as such in the proper coordinate frame because it travels along invariant null-lines.

And with regard to the former concept, it is useful to note Galileo's argument that, to a person riding in the cabin of a moving ship, everything inside the cabin should occur just as if the ship were at rest. It was crucial for Galileo to make this point by \textit{isolating} the inertial system from its relatively moving surroundings---as the point would have been less clear, e.g., if he had argued that when riding in the back of a wagon one can toss a ball straight in the air and have it fall back to the same point within the wagon. However, if one should argue that there \textit{really} can't be privileged observers in the Universe, due to the relativity of inertia, one must go beyond this local-inertial effect---viz.\ the relativity of inertia---and consider the frame with respect to its cosmic surroundings---in which case the argument can't be justified. 

For consider a neutrino, created in a star shortly after the Big Bang: in the neutrino's proper frame, only minutes may have elapsed since it left the star, throughout which time the galaxies would have formed, etc., all moving past it in roughly the same direction, at nearly the speed of light. Clearly the most reasonable interpretation, however, is that the neutrino has \textit{really} been travelling through the Universe for the past 13.8 billion years---and this description may be given, with the cosmic present uniquely and objectively defined, in all frames including the neutrino's. 

Furthermore, if we would assume that there are no privileged observers, it should be noted that the consequence of describing simultaneity and synchronicity as one and the same thing in all frames is a block universe \cite{Putnam1967}---a temporally singular `absolute world' \cite{Minkowski1908} in which `the distinction between past, present, and future has only the significance of a stubborn illusion' \cite{Foelsing1997}; i.e., `The objective world simply \textit{is}, it does not \textit{happen}. Only to the gaze of my consciousness, crawling upward along the life line of my body, does a section of this world come to life as a fleeting image in space which continuously changes in time' \cite{Weyl1949}; `There is no dynamics within space-time itself: nothing ever moves therein; nothing happens; nothing changes. \ldots one does not think of particles ``moving through'' space-time, or as `following along'' their world lines. Rather, particles are just ``in'' space-time, once and for all, and the world line represents, all at once, the complete life history of the particle' \cite{Geroch1978}.

And so I've argued against the simultaneity of synchronicity,---a reasonably intuitive concept held in common between the theories of both Newton and Einstein. But is there any \textit{sensible} justification for the concept that the space in which events \textit{really} take place simultaneously \textit{must} be orthogonal to the proper time-axis of an inertial observer? When our theories are interpreted in this way, is that because one can, e.g., sit down on the floor with legs out in front, raise their right arm out to the side and their left arm up in the air, \textit{and then stick out their tongue in the direction in which time is flowing}, for them as much as it is for their entire surroundings? Of course not. This is no more justified for someone who thus defines a right-handed coordinate system while sitting on solid ground, than it is for a person in the cabin of a ship---whether that is floating on water or flying through space. Therefore, intuition justifies only existence in space that endures with the ticking of everyone's watch---and relativity theory \textit{demands} that this cannot be both coherently defined and synchronous with every inertial observer!

Now, although it may be argued that the alternative assumption of cosmic time is unobservable metaphysics, and therefore unscientific, that simply isn't true---for cosmology does provide strong empirical evidence of an absolute rest-frame in our Universe, as follows. As Einstein noted already in 1917 \cite{Einstein1917}, there appears to be a frame relative to which the bodies of our Universe are at rest, on average. Now, Einstein had no idea of the scope of the Universe at that time, but already by 1923 Weyl realised the significance of this point, which has indeed stood the test of time, when he wrote that \cite{Weyl1923} `Both the papers by de~Sitter \cite{deSitter1917} and Eddington \cite{Eddington1923} lack this assumption on the ``state of rest'' of stars---by the way the only possible one compatible with the homogeneity of space and time. Without such an assumption nothing can be known about the redshift, of course.' For it is true, even in de~Sitter space, that a cosmic time must be assumed in order to calculate redshifts; e.g., for particles in the comoving Lema\^{i}tre-Robertson frame illustrated in Fig.~\ref{fig: dS_LR} and described by Eq.~(\ref{dS_LR}), the redshift will be different from that in the frame of comoving particles in the three-sphere which contracts to a finite radius and subsequently expands (as illustrated by the gridlines of the de~Sitter hyperboloid in Fig.~\ref{fig: dS_LR}) according to
\begin{equation}
\mathrm{d}s^2=-\mathrm{d}T^2+\frac{3}{\Lambda}\cosh^2\left(\sqrt{\frac{\Lambda}{3}}T\right){\mathrm{d}\Omega_3}^2,\label{dS_comoving_univ}
\end{equation} 
where ${\mathrm{d}\Omega_3}$ describes the three-sphere. The existence of more than one formally distinct RW cosmological model in one and the same space-time thus illustrates the importance of defining a cosmic time.

Since 1923, a number of novel observations have strengthened the evidence for a cosmic present, such as Hubble's confirmation of cosmic expansion, the detailed measurement of the expansion rate that has lately been afforded through type Ia supernovae observations, and the discovery of the CMBR, which gives a detailed signature of the cosmic rest-frame relative to which we are in fact moving, according to the common interpretation of its dipole anisotropy. Thus, the assumption of a cosmic present is now very well justified by empirical evidence.

\section{Implications for Cosmology \label{sec_IfC}}
Although many points should be considered in connection to the description of an absolute cosmic present, such as concepts of time travel, free will, and a causally coherent local description of gravitational collapse in the Universe---notwithstanding space-time curvature in general,---the one consequence that I will note pertains to cosmology, and a better explanation of cosmic expansion.

To start, note that in deriving the general line-element for the background geometry of FLRW cosmology, Robertson required four basic assumptions \cite{Rugh2010}: i.\ a congruence of geodesics, ii.\ hypersurface orthogonality, iii.\ homogeneity, and iv.\ isotropy. i.\ and ii.\ are required to satisfy Weyl's postulate of a causal coherence amongst world lines in the entire Universe, by which every single event in the bundle of fundamental world lines is associated with a well-defined three-dimensional set of others with which it `really' occurs simultaneously. However, it seems that ii.\ is therefore mostly required to satisfy the concept that synchronous events in a given inertial frame should have occurred simultaneously, against which I've argued above.

In special relativity, if we allow the fundamental world lines to \textit{set} the cosmic rest-frame, then the cosmic hyperplane should be orthogonal---but that shouldn't be the case in general. Indeed, as I've shown in my PhD thesis \cite{Janzen2012}, in the cosmological Schwarzschild-de~Sitter (SdS) solution,
\begin{equation}
\mathrm{d}s^2=-\frac{r}{\frac{\Lambda}{3}r^3+2M-r}\mathrm{d}r^2+\frac{\frac{\Lambda}{3}r^3+2M-r}{r}\mathrm{d}t^2+r^2\mathrm{d}\Omega^2,\label{SdS_statical}
\end{equation}
for which ${\Lambda}M^2>1/9$, $r>0$ is \textit{timelike}, and $t$ is forever \textit{spacelike}, the $r$-coordinate should well describe the cosmic time \textit{and} factor of expansion in a universe in which, in the coordinates carried by fundamental observers, the cosmic present would not be synchronous, and $r$ would evolve in proper time $\tau$ as
\begin{equation}
r(\tau)\propto\sinh^{2/3}[(\sqrt{3\Lambda}/2)\tau],\label{scalefac}
\end{equation}
which is \textit{incidentally} also the flat $\Lambda$CDM scale-factor of the standard model, that's been empirically constrained this past decade \cite{SNe Ia}; see Appendix~\ref{sec_CSdSCS} for a derivation of Eq.~(\ref{scalefac}) beginning from Eq.~(\ref{SdS_statical}), and a discussion of the result's connection to cosmology. This is the rate of expansion that \textit{all} observers would measure, if distant galaxies were themselves all roughly at rest with respect to fundamental world lines. But in contrast to FLRW theory, this universe actually has to expand---at all $r>0$---as a result of the `de~Sitter effect'; i.e., if such a universe did come to exist at any infinitesimal time, it would \textit{necessarily} expand---and in exactly the manner that we observe---which may be the closest to an explanation of that as we can achieve.

It is, of course, important to stress that this intriguing result is utterly meaningless if simultaneity should rather be defined as synchronicity in a given frame of reference. In that case, as Lema\^{i}tre noted \cite{Lemaitre1949}, the solution describes flat spatial slices extending from $r=0$ to $\infty$, with particles continuously ejected from the origin. It is therefore only by reconceiving the relativistic concepts of time and simultaneity that SdS can be legitimated as a coherent cosmological model with a common origin---and one with the very factor of expansion that we've measured---which really \textit{should} expand, according to the view of expansion as being always driven by $\Lambda$.
\\\\\textbf{Acknowledgements:} Thanks to Craig Callender for reviewing an earlier draft and providing thoughtful feedback that greatly improved this essay. Thanks also to the many participants who commented on and discussed this paper throughout the contest, and FQXi for organising an excellent contest and providing criteria that helped shape the presentation of this argument.

\begin{appendix}
\section{Concerning Schwarzschild-de~Sitter as a Cosmological Solution\label{sec_CSdSCS}}
During the Essay Contest discussions, the critical remarks on this essay that were most important for me, and were by far the most probing, were those offered by George Ellis. Professor Ellis' criticism of the final section indicated, first of all, that the brief mention I made there of a result from my PhD thesis was too underdeveloped to pique much interest in it---and in fact that, stated as it was there, briefly and out of context of the explicit analysis leading from Eq.~(\ref{SdS_statical}) to (\ref{scalefac}), the point was too easily missed. He wrote that the model is `of course spatially inhomogeneous,' when the spatial slices \textit{are} actually homogeneous, but rather are anisotropic; and when I pointed out to him that this is so because, in the cosmological form of the SdS solution $r>0$ is \textit{timelike} and $t$ is forever \textit{spacelike}, he replied that `the coordinate notation is very misleading'. 

So, one purpose of this appendix is to provide the intermediate calculation between Eqs.~(\ref{SdS_statical}) and (\ref{scalefac}), that had to be left out of the original essay due to space limitations---and, in developing a familiarity with the common notation, through the little calculation, to ensure that no confusion remains in regard to the use of $r$ as a timelike variable and $t$ as a spacelike one. For the notation is necessary both in order to be consistent with every other treatment of the SdS metric to date, and because, regardless of whether $r$ is timelike or spacelike in Eq.~(\ref{SdS_statical}), it really does make sense to denote the coordinate with an `$r$' because the space-time is isotropic (i.e. `radially' symmetric) in that direction. 

With these `bookkeeping' items out of the way (after roughly the first four pages), the appendix moves on to address Professor Ellis' two more substantial criticisms, i.e.\ regarding the spatial anisotropy and the fact that the model has no dynamic matter in it; for, as he noted, the model `is interesting geometrically, but it needs supplementation by a dynamic matter and radiation description in order to relate to our cosmic history'. These important points were discussed in the contest forum, but were difficult to adequately address in that setting, so the problem is given more proper treatment in the remaining pages of this appendix once the necessary mathematical results are in-hand. Specifically, in the course of developing a physical picture in which the SdS metric provides the description of a universe that \textit{would} appear isotropic to fundamental observers who measure the same rate of expansion that we do (viz.\ as given by Eq.~(\ref{scalefac})), we will come to a possible, consistent resolution to the problem of accounting for dynamic matter, which leads to a critical examination of the consistency and justification of some of the most cherished assumptions of modern physics, thus further questioning its foundations.

We begin by writing down the equations of motion of `radial' geodesics in the SdS geometry, using them to derive a description of the SdS cosmology that would be appropriate to use from the perspective of fundamental observers who evolve as they do, beginning from a common origin at $r=0$, always essentially \textit{because of} the induced field potential. It will be proved incidentally that the observed cosmological redshifts, in this homogeneous universe which is \textit{not} orthogonal to the bundle of fundamental geodesics---and is therefore precluded by the a priori assumptions of standard FLRW cosmology---must evolve through the course of cosmic time, as a function of the proper time of fundamental observers, with the precise form of the flat $\Lambda$CDM scale-factor---i.e., with exactly the form that has been significantly constrained through observations of type Ia supernovae, baryon acoustic oscillations, and CMBR anisotropies \cite{SNe Ia}.

Since the Lagrangian, 
\begin{equation}
L=-\frac{r-2M-\frac{\Lambda}{3}r^3}{r}\left(\frac{\mathrm{d}t}{\mathrm{d}\tau}\right)^2+\frac{r}{r-2M-\frac{\Lambda}{3}r^3}\left(\frac{\mathrm{d}r}{\mathrm{d}\tau}\right)^2=-1,\label{L_timelike}
\end{equation}
for timelike $(r,t)$-geodesics with proper time $\tau$ in the SdS geometry is independent of $t$, the Euler-Lagrange equations indicate that
\begin{equation}
E\equiv-{1 \over 2}{\partial{L} \over \partial(\mathrm{d}t/\mathrm{d}\tau)}=\frac{r-2M-\frac{\Lambda}{3}r^3}{r}\left(\frac{\mathrm{d}t}{\mathrm{d}\tau}\right)\label{E}
\end{equation}
is conserved ($-2\mathrm{d}E/\mathrm{d}\tau\equiv\mathrm{d}L/\mathrm{d}t=0$). Substituting Eq.~(\ref{E}) into Eq.~(\ref{L_timelike}), then, we find the corresponding equation of motion in $r$:
\begin{equation}
\left(\frac{\mathrm{d}r}{\mathrm{d}\tau}\right)^2=E^2-\frac{r-2M-\frac{\Lambda}{3}r^3}{r}.\label{rtau}
\end{equation}

While the value of $E$ may be arbitrary, we want a value that distinguishes a particular set of geodesics as those describing particles that are `fundamentally at rest'---i.e., we'll distinguish a preferred fundamental rest frame by choosing a particular value of $E$ that meets certain physical requirements. In order to determine which value to use, we first note that where $r$ \textit{is} spacelike, Eq.~(\ref{rtau}) describes the specific (i.e., per unit rest-mass) kinetic energy of a test-particle, as the difference between its (conserved) \textit{specific energy} and the gravitational field's \textit{effective potential},
\begin{equation}
V_{\mathrm{eff}}(r)\equiv\frac{r-2M-\frac{\Lambda}{3}r^3}{r}.\label{V_eff}
\end{equation} 
Then, a reasonable definition sets the `fundamental frame' as the one in which the movement of particles in $r$ and $t$ is essentially \textit{caused} by the non-trivial field potential---i.e., so that $\mathrm{d}r/\mathrm{d}\tau=0$ just where the gravitational potential is identically trivial ($V_{\mathrm{eff}}\equiv1$), and the line-element, Eq.~(\ref{SdS_statical}), reduces to that of Minkowski space. From Eq.~(\ref{rtau}), this amounts to setting $E^2=1$; therefore, a value $E^2>1$ corresponds to a particle that would not come to rest at $r=-\sqrt[3]{6M/\Lambda}$, where $V_{\mathrm{eff}}\equiv1$, but has momentum in $r$ beyond that which would be imparted purely by the field.

As a check that the value $E^2=1$ is consistent with our aims, we can consider its physical meaning another way. First of all, note that where $V_{\mathrm{eff}}\equiv1$, at $r=-\sqrt[3]{6M/\Lambda}$, $r$ is spacelike and $t$ is timelike regardless of the values of $M$ and $\Lambda$; therefore, it is consistent in any case to say that a particle with $E^2>1$ has non-vanishing spatial momentum there. Indeed, from Eq.~(\ref{E}), we find that $t=\tau$ at $r=-\sqrt[3]{6M/\Lambda}$ if, and only if, $E=1$---so the sign of $E$ should in fact be positive for a particle whose proper time increases with increasing $t$ in the absence of gravity. 

Furthermore, note that when $\Lambda=0$, Eq.~(\ref{rtau}) reduces to
\begin{equation}
\left(\frac{\mathrm{d}r}{\mathrm{d}\tau}\right)^2=E^2-1+\frac{2M}{r}.\label{rtau_Sch}
\end{equation}
As such, Misner, Thorne and Wheeler describe $E$ as a particle's `specific energy at infinity', where the effective potential is trivial \cite{Misner1973}. It is relevant to note their statement (on p. 658), that the conservation of 4-momentum `allows and forces one to take over the [term] $E=$``energy at infinity''\ldots, valid for orbits that do reach to infinity, for an orbit that does not reach to infinity.' More generally, we should describe $E$ for arbitrary $M$ and $\Lambda$ as the `energy at vanishing field potential' even when $r=-\sqrt[3]{6M/\Lambda}$ is a negative mathematical abstraction that lies beyond a singularity at $r=0$. In particular, we take $E=1$ to be the \textit{specific energy of a test-particle that is at rest with respect to the vanishing of the potential}. It's simply a matter of algebraic consistency.

Thus, we have $E=1$ as the specific energy of particles that would come to rest in the absence of a gravitational field, which are therefore guided purely through the effective field potential. We therefore use the geodesics with $E=1$ to define a preferred rest frame in the SdS geometries, and say that any particle whose world line is a geodesic with $E\neq1$ is one that has uniform momentum relative to the fundamental rest frame.

We can now write the SdS line-element, Eq.~(\ref{SdS_statical}), in the proper frame of a bundle of these fundamental geodesics, which evolve through $t$ and $r$ all with the same proper time, $\tau$, and occupy constant positions in `space'. Since $\Lambda$ must be positive in order to satisfy the requirement, ${\Lambda}M^2>1/9$, for $r>0$ to be timelike---i.e.\ the requirement for the SdS line-element to be cosmological rather than a local solution---it is more convenient to work with scale-invariant parameters $r\rightarrow{r'}=\sqrt{\frac{\Lambda}{3}}r$, $t\rightarrow{t'}=\sqrt{\frac{\Lambda}{3}}t$, $\tau\rightarrow{\tau'}=\sqrt{\frac{\Lambda}{3}}\tau$, $M\rightarrow{M'}=\sqrt{\frac{\Lambda}{3}}M$, etc., normalising all dimensional quantities by the cosmic length-scale $\sqrt{3/\Lambda}$ (see, e.g., \S~{66} in \cite{Eddington1923} or \S~{4} in \cite{Dyson1972} for interesting discussions of this length parameter). This normalisation ultimately amounts to striking out the factor $\Lambda/3$ from the $r^3$-term in the line-element $\left(\mathrm{since}\ \frac{\sqrt{\frac{\Lambda}{3}}r-2\sqrt{\frac{\Lambda}{3}}M-\left(\sqrt{\frac{\Lambda}{3}}r\right)^3}{\sqrt{\frac{\Lambda}{3}}r}\rightarrow\frac{r-2M-r^3}{r}\right)$, or e.g.\ writing the flat $\Lambda$CDM scale-factor, Eq.~(\ref{scalefac}), as
\begin{equation}
r(\tau)\propto\sinh^{2/3}\left(3\tau/2\right),\label{scalefac_scaleinvtau}
\end{equation}
and the corresponding Hubble parameter as
\begin{equation}
H\equiv\dot{r}/r=\coth(3\tau/2),\label{Hubble}
\end{equation}
which exponentially approaches $H=1$ on timescales $\tau\sim2/3$.

The evolution of each geodesic through scale-invariant $t$ and $r$ is then given, through Eqs.~(\ref{E}) and (\ref{rtau}) with $E=1$, as 
\begin{equation}
\partial_{\tau}{t}\equiv\frac{\partial{t}}{\partial\tau}=\frac{r}{r-2M-r^3},\label{tdot}
\end{equation}
\begin{equation}
(\partial_{\tau}r)^2\equiv\left(\frac{\partial{r}}{\partial\tau}\right)^2=\frac{2M+r^3}{r}.\label{rdot}
\end{equation} 
Eq.~(\ref{rdot}) can be solved using $\int\left(u^2+a\right)^{-1/2}\mathrm{d}u=\ln\left(u+\sqrt{u^2+a}\right)$ after substituting $u^2=2M+r^3$. Taking the positive root (so $\tau$ increases with $r$), we have,
\begin{equation}
\tau=\int^{r(\tau)}_{r(0)}\sqrt{\frac{r}{2M+r^3}}\mathrm{d}r=\left.\frac{2}{3}\ln\left(\sqrt{2M+r^3}+r^{3/2}\right)\right|^{r(\tau)}_{r(0)},\label{t_int}
\end{equation}
where the lower limit on $\tau$ has been arbitrarily set to 0. Thus, in this frame we can express $r$ as a function of each observer's proper time $\tau$ and an orthogonal (i.e. synchronous, with constant $\tau=0$) spatial coordinate, $r(0)$, which may be arbitrarily rescaled without altering the description in any significant way.

Then, as long as $M$ is nonzero, a convenient set of coordinates from which to proceed results from rescaling the spatial coordinate as\footnote{Note that this transformation is not valid when $M=0$, which we are anyhow not interested in. An equivalent transformation in that case is found by setting $r(0)\equiv{e}^{\chi}$, whence $r(\tau,\chi)=e^{\tau+\chi}$, and Eqs.~(\ref{g_chit}), (\ref{g_tt}), and (\ref{g_chichi}), below, yield the line-element, $\mathrm{d}s^2=-\mathrm{d}\tau^2+r^2\left(\mathrm{d}\chi^2+\mathrm{d}\theta^2+\sin^2\theta{\mathrm{d}\phi^2}\right)$.} 
\begin{equation}
r(0)\equiv(2M)^{1/3}\sinh^{2/3}\left(\frac{3}{2}\chi\right) ;~M\neq0, \label{r(0)}
\end{equation}
from which we find, after some rearranging of Eq.~(\ref{t_int}),\footnote{Note that the two identities, $e^x=\sinh(x)+\cosh(x)$ and $\mathrm{arsinh}(x)=\ln\left(x+\sqrt{x^2+1}\right)$, are useful here. Eq.~(\ref{r(tau,chi)}), along with our eventual line-element, Eq.~(\ref{SdS_proper}), was originally found by Lema\^{i}tre \cite{Lemaitre1949}, although his solution to Eq.~(\ref{rdot}) (with dimensionality restored),
\[r=(6M/\Lambda)^{1/3}\sinh^{2/3}\left[3\sqrt{\Lambda}(t-t_0)/2\right],\]
is too large in its argument by a factor of $\sqrt{3}$.}
\begin{equation}
r(\tau,\chi)=(2M)^{1/3}\sinh^{2/3}\left(\frac{3}{2}[\tau+\chi]\right),\label{r(tau,chi)}
\end{equation}
which immediately shows the usefulness of rescaling the $r(0)$ as in Eq.~(\ref{r(0)}), since it allows Eq.~(\ref{t_int}) to be solved explicitly for $r(\tau,\chi)$. As such, we immediately have the useful result (cf.\ Eq.~(\ref{rtau})),
\begin{equation}
\partial_{\chi}r\equiv\frac{\partial{r}}{\partial\chi}=\partial_{\tau}r=\sqrt{\frac{2M+r^3}{r}}.\label{rprime}
\end{equation}

The transformation, $t(\tau,\chi)$, may then be calculated from
\begin{equation}
\frac{\mathrm{d}t}{\mathrm{d}r}=\frac{\partial_{\tau}t}{\partial_{\tau}r}=\frac{r}{r-2M-r^3}\sqrt{\frac{r}{2M+r^3}}.\label{dtdr}
\end{equation}
Then, to solve for $t(\tau,\chi)$, we can gauge the lower limits of the integrals over $t$ and $r$, at $\tau=0$, by requiring that their difference, defined by 
\begin{equation}
t(\tau,\chi)=\int^{r(\tau,\chi)}\frac{r}{r-2M-r^3}\sqrt{\frac{r}{2M+r^3}}\mathrm{d}r-F(\chi),\label{t(tau,chi)}
\end{equation}
sets 
\begin{equation}
0=g_{\chi\tau}=g_{tt}\partial_{\tau}t\partial_{\chi}t+g_{rr}\partial_{\tau}r\partial_{\chi}r.\label{g_chit}
\end{equation} 
(Thus, $\chi$ will be orthogonal to $\tau$.) This calculation is straightforward:\footnote{Note that we don't actually have to solve the integral in Eq.~(\ref{t(tau,chi)}), since only partial derivatives of $t$ are needed here and below.}
\begin{eqnarray}
0\!\!\!&=&\!\!\!-\frac{r}{r-2M-r^3}+\partial_{\chi}F(\chi)+\frac{r}{r-2M-r^3}\frac{2M+r^3}{r}\\
\!\!\!&=&\!\!\!\partial_{\chi}F(\chi)-1,
\end{eqnarray}
so that $F(\chi)=\chi$.

Now, it is a simple matter to work out the remaining metric components as follows: our choice of proper reference frame immediately requires 
\begin{equation}
g_{\tau\tau}=g_{tt}(\partial_{\tau}t)^2+g_{rr}(\partial_{\tau}r)^2=-1,\label{g_tt}
\end{equation} 
according to the Lagrangian, Eq.~(\ref{L_timelike}); and by direct calculation, we find
\begin{eqnarray}
g_{\chi\chi}\!\!\!&=&\!\!\!g_{tt}(\partial_{\chi}t)^2+g_{rr}(\partial_{\chi}r)^2\\
\!\!\!&=&\!\!\!-\frac{r-2M-r^3}{r}\left(\frac{2M+r^3}{r-2M-r^3}\right)^{2}+\frac{r}{r-2M-r^3}\frac{2M+r^3}{r}\\
\!\!\!&=&\!\!\!(\partial_{\chi}r)^2.\label{g_chichi}
\end{eqnarray} 
But this result is independent of any arbitrary rescaling of $\chi$; for if we replaced $\chi=f(\xi)$ in Eq.~(\ref{r(0)}), we would then find the metric to transform as $g_{\xi\xi}=g_{\chi\chi}(d\chi/d\xi)^2=(\partial_{\xi}r)^2$, the other components remaining the same. 

Therefore, the SdS metric in the proper frame of an observer who is cosmically `at rest', in which the spatial coordinates are required, according to an appropriate definition of $F(\chi)$, to be orthogonal to $\tau$,\footnote{Note that the `radially' symmetric part of Eq.~(\ref{SdS_statical}) is already orthogonal to $\tau$.} can generally be written,
\begin{equation}
\mathrm{d}s^2=-\mathrm{d}\tau^2+\left(\partial_{\chi}r\right)^2\mathrm{d}\chi^2+r^2\left(\mathrm{d}\theta^2+\sin^2\theta{\mathrm{d}\phi}^2\right).\label{SdS_proper}
\end{equation}
This proves Lema\^{i}tre's result from 1949 \cite{Lemaitre1949},---that slices $\mathrm{d}\tau=0\left(\Rightarrow\left(\partial_{\chi}r\right)^2\mathrm{d}\chi^2=\left(\mathrm{d}r/\mathrm{d}\chi\right)^2\mathrm{d}\chi^2=\mathrm{d}r^2\right)$ are Euclidean, with line-element,
\begin{equation}
\mathrm{d}\sigma^2=\mathrm{d}r^2+r^2\left(\mathrm{d}\theta^2+\sin^2\theta{\mathrm{d}\phi}^2\right).
\end{equation}
However, in the course of our derivation we have also found that Lema\^{i}tre's \textit{physical} interpretation---that the `geometry is Euclidean on the expanding set of particles which are ejected from the point singularity at the origin'---is wrong. 

It is \textit{wrong} to interpret this solution as describing the evolution of synchronous `space' which always extends from $r=0$ to $r=+\infty$ along lines of constant $\tau$, being truncated at the $r=0$ singularity at $\chi=-\tau$ from which particles are continuously ejected as $\tau$ increases. But this is exactly the interpretation one is apt to make, who is accustomed to thinking of synchronous spacelike hypersurfaces as `space' that exists `out there', regardless of the space-time geometry or the particular coordinate system used describe it.

As we noted from the outset, the `radial' geodesics that we have now described by the lines $\chi=const.$, along which particles all measure their own proper time to increase with $\tau$, describe the world lines of particles that are all fundamentally at rest---i.e., at rest with respect to the vanishing of the effective field potential. Therefore, these particles should not all emerge from the origin at different times, and then somehow evolve together as a coherent set; but \textit{by Weyl's principle} they should all emerge from a \textit{common origin}, and evolve through the field that varies \textit{isotropically in} $r$, \textit{together for all time}. In that case, space will be homogeneous, since the constant cosmic time ($\mathrm{d}r=0$) slices of the metric can be written independent of spatial coordinates; so every fundamental observer can arbitrarily set its spatial position as $\chi=0$ and therefore its origin in time as $\tau=0$.

The spaces of constant cosmic time should therefore be those slices for which $r(\tau+\chi)=const.$---i.e., we set $\bar{\tau}=\tau+\chi$ as the proper measure of cosmic time in the fundamental rest frame of the universe defined by this coherent bundle of geodesics, so that Eq.~(\ref{r(tau,chi)}) becomes
\begin{equation}
r(\bar{\tau})=(2M)^{1/3}\sinh^{2/3}\left(3\bar{\tau}/2\right).\label{SdS_scalefac}
\end{equation}
The spacelike slices of constant $\bar{\tau}$ are at $45^{\circ}$ angles in the $(\tau,\chi)$-plane, and are therefore definitely not synchronous with respect to the fundamental geodesics. However, given this definition of cosmic time, the redshift of light that was emitted at $\bar{\tau}_e$ and is observed now, at $\bar{\tau}_0$, should be
\begin{equation}
1+z=\frac{r(\bar\tau_0)}{r(\bar\tau_e)},\label{z_SdS}
\end{equation}
where $r(\bar{\tau})$ has exactly the form of the flat $\Lambda$CDM scale-factor (cf. Eq. (\ref{scalefac_scaleinvtau})), which is exactly the form of expansion in our Universe that has been increasingly constrained over the last fifteen years \cite{SNe Ia}.\\\\\\
Now, in order to properly theoretically interpret this result for the observed redshift in our SdS cosmology, it should be considered in relation to FLRW cosmology---and particularly the theory's basic assumptions. As noted in \S~\ref{sec_IfC}, the kinematical assumptions used to constrain the form of the line-element are: i.\ a congruence of geodesics, ii.\ hypersurface orthogonality, iii.\ homogeneity, and iv.\ isotropy. Assumptions i.\ and ii.\ have a lot to do with how one defines `simultaneity', which I have discussed both in the context of special relativity in \S~\ref{sec_TCP}, and now in the context of the SdS cosmology, in which simultaneous events that occur in the course of cosmic time are \textit{not} synchronous \textit{even in the fundamental rest frame}. As the discussion should indicate, the definition of `simultaneity' is somewhat arbitrary---and it is an \textit{assumption} in any case---and should be made with the physics in mind. Einstein obviously had the physics in mind when he proposed using an operational definition of simultaneity \cite{Einstein1905}; but it has since been realised that even special relativity, given this definition, comes to mean that time can't pass, etc., as noted in \S~\ref{sec_TCP}. 

Special relativity \textit{should} therefore be taken as an advance on Newton's bucket argument, indicating that not only should acceleration be absolute, as Newton showed (see, e.g., \cite{Maudlin2012} for a recent discussion of Newton's argument), but velocity should be as well, since time obviously passes---which it can't do, according to special relativity, if motion isn't \textit{always} absolute. Usually, however, the opposite is done, and people who have been unwilling or unable to update the subjective and arbitrary definitions of simultaneity, etc., from those laid down by Einstein in 1905, have simply concluded that Physical Reality has to be a four-dimensional Block in which time doesn't pass, and the apparent passage of time is a stubborn illusion; see, e.g., \S~5 in \cite{Greene2004} for a popular account of this, in addition to Refs.~\cite{Putnam1967,Minkowski1908,Foelsing1997,Weyl1949,Geroch1978}. The discussion in \S~\ref{sec_TCP} shows how to move forward with a realistic, physical, and most importantly a \textit{relativistic} description of objective temporal passage, which can be done only when `simultaneity' is not equated with `synchronicity' a priori; and another useful thought-experiment along those lines, which shows how perfectly acceptable it is to assume objective temporal passage in spite of relativistic effects, is presented in my more recent FQXi essay \cite{Janzen2013}.

In contrast to the hardcore relativists who would give up temporal passage in favour of an operational definition of simultaneity, Einstein was the first relativist to renege on truly relative simultaneity when he assumed an absolute time in constructing his cosmological model \cite{Einstein1917}; and despite immediately being chastised by de~Sitter over this \cite{deSitter1917a}, he never did balk in making the same assumption whenever he considered the cosmological problem \cite{Einstein1931b,Einstein1932,Einstein1945}---as did just about every other cosmologist who followed, with very few notable exceptions (e.g., de Sitter \cite{deSitter1917a,deSitter1917} was one, as there was no absolute time implicit in his model). 

But whenever the assumption of absolute time has been made in cosmology, it has been made together with special relativity's baggage, as the slices of true simultaneity have been assumed to be synchronous in the fundamental rest frame. Now we see that, not only is the operational definition \textit{wrong} in the case of special relativity (since it comes to require that time does not pass, which is realistically unacceptable), but here we've considered a general relativistic example in which equating `simultaneity' and `synchronicity' makes even less sense in terms of a reasonable physical interpretation of the mathematical description, since the interpretation is \textit{causally incoherent}---i.e. Lema\^{i}tre's interpretation, that the line-element Eq.~(\ref{SdS_proper}) should describe an `expanding set of particles which are ejected from the point singularity at the origin' represents \textit{abominable physical insight}. The main argument of this essay was therefore, that while assumption i.\ of FLRW cosmology is justified from the point of view that relative temporal passage should be coherent, assumption ii.\ is not, and this unjustified special relativistic baggage should be shed by cosmologists---and really by all relativists, as it leads to further wrong interpretations of the physics.

Assumption iii.\ hardly requires discussion. It is a mathematical statement of the cosmological principle---that no observer holds a special place, but the Universe should look the same from every location---and is therefore as fundamental an assumption as the principle of relativity. Furthermore, our SdS universe \textit{is} homogeneous, so there is no problem.

The final assumption, however, is a concern. The isotopy of our Universe is an empirical fact---it looks the same to us in every direction, and the evidence is that it must have done since its beginning. In contrast, the spatial slices of the cosmological SdS solution are \textit{not} isotropic: they are a 2-sphere with extrinsic radius of curvature $r$, multiplied by another dimension that scales differently with $r$. Furthermore, by Eq.~(\ref{dtdr}) we know that all these fundamental world lines move uniformly through this third spatial dimension, $t$, as $r$ increases. 

The SdS cosmology therefore describes a universe that should be conceived as follows: First of all suppressing one spatial dimension, the universe can be thought of as a 2-sphere that expands from a point, with all fundamental observers forever motionless along the surface; then, the third spatial dimension should be thought of as a line at each point on the sphere, through which fundamental observers travel uniformly in the course of cosmic time.

Since it is a general relativistic solution, the distinction between curvature along that third dimension of space and motion through it is not well-defined. However, a possibility presents itself through an analogy with the \textit{local} form of the SdS solution. As with all physically meaningful solutions of the Einstein field equations, this one begins with a physical concept from which a general line-element is written; then the line-element gets sent through the field equations and certain restrictions on functions of the field-variables emerge, allowing us to constrain the general form to something more specific that satisfies the requisite second-order differential equations. This is, e.g., also how FLRW cosmology is done---i.e., first the RW line-element satisfying four basic physical/geometrical requirements is written down, and then it is sent through the field equations to determine equations that restrict the form of the scale-factor's evolution, under a further assumption that finite matter densities in the universe should influence the expansion rate. The local SdS solution, too, is derived as the vacuum field that forever has spacelike radial symmetry about some central gravitating body, and the field equations are solved to restrict the form of the metric coefficients in the assumed coordinate system. But then, as Eddington noted \cite{Eddington1923},
\begin{quote}
We reach the same result if we attempt to define symmetry by the propagation of light, so that the cone $ds=0$ is taken as the standard of symmetry. It is clear that if the locus $ds=0$ has complete symmetry about an axis (taken as the axis of $t$) $ds^2$ must be expressable by [the radially isotropic line-element with general functions for the metric coefficients].
\end{quote}
Therefore, the local SdS metric corresponds to the situation in which light propagates isotropically, and its path in space-time is described by the null lines of a Lorentzian metric. Prior to algebraic abstraction (i.e. the assumption of a Lorentzian metric and a particular coordinate system), the geometrical picture is already set; and it is upon that basic geometrical set-up that the algebraic properties of the general relativistic field are imposed. 

This construction of the local SdS solution through physical considerations of light-propagation can be used analogously in constructing a geometrical picture upon which the cosmological SdS solution can be based; 
however, some more remarks are necessary before coming to that. First of all, as our discussion of the local SdS and FLRW solutions indicates, in general much of the \textit{physics} enters into the mathematical description already in defining the basic geometrical picture and the corresponding line-element, which broadly sets-up the physical situation of interest. Only then is the basic physical picture further constrained by requiring that it satisfy the specific properties imposed by Einstein's field law. In fact, when it comes to the cosmological problem, and we begin \textit{as always} by assuming what will be our \textit{actual} space, and how it will roughly evolve as \textit{actual} cosmic time passes---i.e. by \textit{first} assuming prior \textit{kinematical} definitions of absolute space and time, and then constructing an appropriate cosmological line-element, which is finally constrained through the \textit{dynamical} field law---there is really a lot of room to make it as we like it.

But now we have a particular line-element in mind (viz.\ the SdS cosmological solution), and we can use it in guiding our basic kinematical definitions. In particular, we have the description of a universe that \textit{is} a two-dimensional sphere that expands as a well-defined function of the proper time of fundamental observers who all remain absolutely at rest, multiplied by another dimension through which those same observers \textit{are} moving, uniformly at a rate that varies through the course of cosmic time. According to the equivalence principle, it may be that the gravitational field is non-trivial along that particular direction of space, and therefore guides these fundamental observers along---or it could be that this direction is uniform as well, and that the fundamental observers are moving along it, and therefore describe it relatively differently. 

What is interesting about this latter possibility, is that there would be a fundamental background metric describing the evolution of this uniform space, and that the metric used by fundamental observers to describe the evolution of space-time in their proper frames shouldn't necessarily \textit{have} to be the same fundamental metric transformed to an appropriate coordinate system. The metric itself might be defined differently from the background metric, for other physical reasons. In this case, an affine connection defining those world lines as fundamental geodesics may not be \textit{compatible} with the more basic metric, and could be taken as the covariant derivative of a different one. The picture starts to resemble teleparallelism much more closely than it does general relativity; but since the two theories are equivalent, and we have recognised that in any case the kinematical definitions must be made first---i.e.\ since we must set-up the kinematical definitions in the first place, according to the physical situation we want to describe, before ensuring that the resulting line-element satisfies the field equations---we'll press on in this vein.

Let us suppose a situation where there is actually no gravitational mass at all, but fundamental inertial observers---the constituent \textit{dynamical matter} of our system---are \textit{really} moving uniformly through a universe that fundamentally \textit{is} isotropic and homogeneous, and expands through the course of cosmic time. The fundamental metric for this universe should satisfy even the RW line-element's orthogonality assumption, although the slices of constant cosmic time would \textit{not} be synchronous in the rest frame of the fundamental observers. Since space, in the two-dimensional slice of the SdS cosmology through which fundamental observers are \textit{not} moving, really \textit{is} spherical, the obvious choice is an expanding 3-sphere, with line-element
\begin{equation}
\mathrm{d}s^2=-\mathrm{d}T^2+R(T)^2{\mathrm{d}\Omega_3}^2,\label{3sphere}
\end{equation} 
where the radius $R(T)$ varies, according to the vacuum field equations, as
\begin{equation}
R(T) = C_1 e^{\sqrt{\Lambda \over 3}T}+C_2 e^{-\sqrt{\Lambda \over 3}T};\ C_1C_2=\frac{3}{4\Lambda}.
\end{equation}
In particular, because a teleparallel theory would require a parallelisable manifold, we note that this is true if
\begin{equation}
C_1=C_2=\frac{1}{2}\sqrt{\frac{3}{\Lambda}},
\end{equation}
and the de~Sitter metric is recovered (cf.~Eq.~(\ref{dS_comoving_univ})). While there may be concern because in this case $R(0)=\sqrt{3/\Lambda}>0$ is a minimum of contracting and expanding space, I will argue below that this may actually be an advantage.

This foliation of de~Sitter space is particularly promising for a couple of reasons: i.\ the bundle of fundamental geodesics in Eq.~(\ref{dS_comoving_univ}) are world lines of massless particles, i.e.\ the ones at $r=0$ for all $t$ in Eq.~(\ref{SdS_statical}) with $M=0$; and ii.\ unlike e.g. the 2-sphere (or spheres of just about every dimension), the 3-sphere \textit{is} parallelisable, so it is possible to define an objective direction of motion for the dynamic matter.


Now we are finally prepared to make use, by analogy, of Eddington's remark on the derivation of the local SdS metric in terms of light propagation along null lines. In contrast, we are now beginning from the description of a universe in which particles that \textit{are not} moving through space are massless, but we want to write down a different Lorentzian metric to describe the situation from the perspective of particles that \textit{are} all moving through it at a certain rate, who define null lines as the paths of relatively moving masselss particles---so, we will write down a new metric to use from the perspective of particles that all move \textit{along null lines} pointing in one direction of de Sitter space, describing the relatively moving paths of massless particles that actually remain motionless in the 3-sphere, as null lines instead. This new line-element has the form,
\begin{equation}
\mathrm{d}s^2=-A(r,t)\mathrm{d}r^2+B(r,t)\mathrm{d}t^2+r^2\mathrm{d}\Omega^2,
\end{equation}
where $r$ points in the \textit{timelike} direction of the universe's increasing radius, and $t$ describes the dimension of space through which the fundamental particles are moving. Solving Einstein's field equations proves the Jebsen-Birkhoff theorem---that $A$ and $B$ are independent of the spacelike variable $t$---and leads to the cosmological SdS solution, Eq.~(\ref{SdS_statical}), as the abstract description of this physical picture.

Thus, we have come full circle to a statement of the line-element that we started with. Our analysis began with a proof that in this homogeneous universe, redshifts should evolve with exactly the form that they do in a flat $\Lambda$CDM universe; and in the last few pages we have aimed at describing a physical situation in which this line-element would apply in the proper reference frame of dynamical matter, and the observed large scale structure would be isotropic. And indeed, in this universe, in which the spatial anisotropy in the line-element is an artifact of the motion of fundamental observers through homogeneous and isotropic expanding space \textit{and} their definition of space-time's null lines, this would be so---for as long as these fundamental observers are uniformly distributed in space that really \textit{is} isotropic and homogeneous, all snapshots of constant cosmic time (and therefore the development, looking back in time with increasing radius all the way to the cosmological horizon) should appear isotropic, since these uniformly distributed observers would always be at rest \textit{relative to one another}.\\\\\\
Having succeeded in showing how the SdS metric can be used to describe a homogeneous universe in which the distribution of dynamical matter appears isotropic from each point, and which would be measured to expand at exactly the rate described by the flat $\Lambda$CDM scale-factor---i.e.\ with exactly the form that has been observationally constrained \cite{SNe Ia}, but which has created a number of problems because, from a basic theoretical standpoint, many aspects of the model are not what we expect---we should conclude with a brief discussion of potential implications emerging from the hypothesis that the SdS cosmology accounts for the fundamental background structure in our Universe.

The most obvious point to note is that the hypothesis would challenge what Misner, Thorne, and Wheeler called `Einstein's explanation for ``gravitation"'---that `\textit{Space acts on matter, telling it how to move. In turn, matter reacts back on space, telling it how to curve}' \cite{Misner1973}. For according to the SdS cosmology, the structure of the absolute background should remain unaffected by its matter-content, and only the geometry of space-time---the four-dimensional set of events that develops as things happen in the course of cosmic time---will depend on the locally-inertial frame of reference in which it is described.

When I mentioned to Professor Ellis in the essay contest discussions, that I think the results I have now presented in this appendix may provide some cause to seriously reconsider the assumption that the expansion rate of the Universe should be influenced by its matter-content---a fundamental assumption of standard cosmology, based on `Einstein's explanation for ``gravitation"', which Professor Loeb also challenged in his submission to the contest---his response was, `Well its not only of cosmology its gravitational theory. It describes solar system dynamics, structure formation, black holes and their interactions, and gravitational waves. The assumption is that the gravitational dynamics that holds on small scales also holds on large scales. It's worked so far.' And indeed, it has worked so far---but that is not a good reason to deny consideration of alternate hypotheses. In fact, as noted by Einstein in the quotation that began this essay, `Concepts that have proven useful in the order of things, easily attain such an authority over us that we forget their Earthly origins and accept them as unalterable facts\ldots The path of scientific advance is often made impassable for a long time through such errors.' From Einstein's point of view, such a position as `It's worked so far' is precisely what becomes the greatest barrier to scientific advance. 

After making this point, Einstein went on to argue that it is really when we challenge ourselves to rework the basic concepts we have of Nature that fundamental advances are made, adding \cite{Einstein1916},
\begin{quote}
This type of analysis appears to the scholars, whose gaze is directed more at the particulars, most superfluous, splayed, and at times even ridiculous. The situation changes, however, when one of the habitually used concepts should be replaced by a sharper one, because the development of the science in question demanded it. Then, those who are faced with the fact that their own concepts do not proceed cleanly raise energetic protest and complain of revolutionary threats to their most sacred possessions.
\end{quote}

In this same spirit, we should note that not only our lack of a technical reason for why the Universe should ever have come to expand---i.e.\ the great explanatory deficit of modern cosmology described in \S~\ref{sec_OCE}---but also both the cosmological constant problem \cite{Weinberg1989} and the horizon problem \cite{Guth1981} are significant problems under the assumption that cosmic expansion is determined by the Universe's matter-content. The former of these two is the problem that the vacuum does not appear to gravitate: an optimistic estimate, that would account for only the electron's contribution to the vacuum energy, still puts the theoretical value $10^{30}$ larger than the dark energy component that cosmologists have measured experimentally \cite{Burgess2013}. The latter problem is that we should not expect the observable part of a general relativistic big bang universe to be isotropic, since almost all of what can now be seen would not have been in causal contact prior to structure formation---and indeed, the light reaching us now from antipodal points of our cosmic event horizon has come from regions that \textit{still} remain out of causal contact with each other, since they are only just becoming visible to \textit{us}, at the half-way point between them.

While there is no accepted solution to the cosmological constant problem, the horizon problem is supposed to be resolved by the inflation scenario \cite{Guth1981}---an epoch of exponential cosmic expansion, proposed to have taken place almost instantly after the Big Bang, which would have carried off our cosmic horizon much faster than the speed of light, leaving it at such a great distance that it is only now coming back into view. Recently, provisional detection of B-mode polarisation in the CMBR \cite{Ade2014} that is consistent with the theory of gravitational waves from inflation \cite{IGW} has been widely lauded as a `proof' of the theory. However, in order to reconcile apparent discrepancies with measurements of the CMBR's anisotropy signature from the \textit{Planck} satellite, BICEP2 researchers have suggested that ad hoc tweaks of the $\Lambda$CDM model may be necessary \cite{Commissariat2014}. 

Details involving the emergence of dynamical matter in the SdS cosmology have not been worked out; however, there is no reason to suspect ab initio that the gravitational waves whose signature has potentially been preserved in the CMBR, would not also be produced in the scenario described here. More importantly, though, the SdS cosmology provides a description that precisely agrees with the observed large-scale expansion of our Universe, and does so without the need to invent any ad hoc hypotheses in order to `save the appearances' that we have found to be so very different from our prior theoretical expectations. The theory simultaneously \textit{solves} the expansion problem outlined in \S~\ref{sec_OCE} (since expansion \textit{must} proceed at an absolute rate, regardless of the universe's dynamical matter-content) and \textit{subverts} the major issues associated with the assumption that cosmic expansion is determined by the Universe's matter-content.

In the SdS cosmology, the vacuum \textit{can} very well gravitate locally without affecting the cosmic expansion rate, and the universe \textit{should} appear, from every point, to be isotropic out to the event horizon. Even the flatness problem---viz.\ the problem that the curvature parameter has been constrained very precisely around zero, when the expectation, again under the assumption that the Universe's structure should be determined by its matter-content, is that it should have evolved to be very far from that---which inflation is also meant to resolve, is subverted in this picture. For indeed, the universe described here, despite being closed, would be described by fundamental observers to expand exactly  according to the \textit{flat} $\Lambda$CDM scale-factor, Eq.~(\ref{SdS_scalefac}).

Additionally, the SdS cosmology provides a fundamentally different perspective on the so-called `coincidence problem'---viz.\ that the matter and dark energy density parameters of the standard FLRW model are nearly equal, when for all our knowledge they could easily be hundreds of orders of magnitude different. Indeed, if we write down the Friedman equation for a flat $\Lambda$CDM universe using Eq.~(\ref{Hubble}) with dimensionality restored,
\begin{equation}
H^2=\frac{\Lambda}{3}\coth^2\left(\frac{\sqrt{3\Lambda}}{2}\tau\right)=\frac{1}{3}(8\pi\rho+\Lambda),
\end{equation}
we find that the matter-to-dark energy density ratio, $\epsilon\equiv8\pi\rho/\Lambda$, while infinite at the big bang, should approach zero exponentially quickly. For example, given the measured value $\Lambda\sim10^{-35}~\mathrm{s}^{-2}$, we can write
\begin{equation}
\tau=\frac{2}{\sqrt{3\Lambda}}\mathrm{arcoth}\,\sqrt{\epsilon+1}\approx11~\mathrm{Gyr}\cdot\mathrm{arcoth}\,\sqrt{\epsilon+1}.
\end{equation}
From here, we find that the dark energy density was $1\%$ of the matter density at $\tau(\epsilon=100)=1$ billion years after the big bang, and that the matter density will be $1\%$ of the dark energy density at $\tau(\epsilon=0.01)=33$ billion years after the big bang.

At first glance, these results may seem to indicate that it is not so remarkable that $\epsilon$ should have a value close to 1 at present. However, from an FLRW perspective, the value of $\Lambda$ could really have been anything---and if it were only $10^4$ larger than it is (which is indeed still far less than our theoretical predictions), $\epsilon$ would have dropped below $1\%$ already at $\tau=0.3$ billion years, and the Universe should now, at 13.8 billion years, be nearly devoid of observable galaxies (so we would have trouble detecting $\Lambda$'s presence). On the other hand, if $\Lambda$ were smaller by $10^4$, it would only become detectable after 100 billion years. Therefore, from an FLRW perspective it is indeed a remarkable coincidence that $\Lambda$ has the particular order of magnitude that it has, which has allowed us to conclusively detect its presence in our Universe.

In contrast to current views, within the FLRW paradigm, on the problem that we know of no good reason why $\Lambda$ should have the particularly special, very small value that it has---which has often led to controversial discussions involving the anthropic principle and a multiverse setting---the SdS cosmology again does not so much `solve' the problem by explaining why the particular value of $\Lambda$ should be observed, but really offers a fundamentally different perspective in which the same problem simply does not exist. For indeed, while the mathematical form of the observed scale-factor in the SdS cosmology should equal that of a flat $\Lambda$CDM universe, the energy densities described in Friedmann's equations are only effective parameters within the former framework, and really are of no essential importance. And in fact, as the analysis in the first part of this Appendix indicates (and again, cf.\ the sections by Eddington \cite{Eddington1923} and Dyson \cite{Dyson1972} referenced above), $\Lambda$ fundamentally sets the scale in the SdS universe: from this perspective, $\sqrt{3/\Lambda}$, which has an empirical value on the order of 10 billion years, \textit{is} the fundamental timescale. It therefore makes little sense to question what effect different values of $\Lambda$ would have on the evolution of an SdS universe, since $\Lambda$ sets the scale of everything a priori; thus, the observable universe should rescale with any change in the value of $\Lambda$. But for the same reason, it is interesting to note that the present order of things has arisen on roughly this characteristic timescale. From this different point-of-view, then, the more relevant question to ponder is: Why should the structure of the subatomic world be such that when particles did coalesce to form atoms and stars, those stars evolved to produce the atoms required to form life-sustaining systems, etc., all on roughly this characteristic timescale?

Despite the SdS cosmology's many attractive features, it may still be objected that the specific geometrical structure of the SdS model entails a significant assumption on the fundamental geometry of physical reality, for which there should be a reason; i.e., the question arises: if the geometry is not determined by the world-matter, then by what? While a detailed answer to this question has not been worked out (although, see \S~3.3 in \cite{Janzen2012}), it is relevant to note that the local form of the SdS solution---which is only parametrically different from the cosmological form upon which our analysis has been based; i.e.\ ${\Lambda}M^2<1/9$ rather than ${\Lambda}M^2>1/9$---is the space-time description outside a spherically symmetric, uncharged black hole, which is exactly the type that is expected to result from the eventual collapse of every massive cluster of galaxies in our Universe---even if it takes all of cosmic time for that collapse to finally occur. In fact, there seems to be particular promise in this direction, given that the singularity at $r=0$ in the SdS cosmology is not a real physical singularity, but the artifact of a derivative metric that must be ill-defined there, since space must actually always have a finite radius according to the fundamental metric, Eq.~(\ref{3sphere}). This is the potential advantage that was noted above, of the finite minimum radius of the foliation of de~Sitter space defined in Eq.~(\ref{dS_comoving_univ}). And as far as that goes, it should be noted that the Penrose-Hawking singularity theorem `cannot be directly applied when a \textit{positive cosmological constant} $\lambda$ is present' \cite{Hawking1970}, which is indeed our case. For all these reasons, we might realistically expect a description in which gravitational collapse leads to universal birth, and thus an explanation of the Big Bang and the basic cosmic structure we've had to assume.

Along with such new possibilities as an updated description of collapse, and of gravitation in general, that may be explored in a relativistic context when the absolute background structure of cosmology is objectively assumed, the SdS cosmology, through its specific requirement that the observed rate of expansion \textit{should be} described exactly by the flat $\Lambda$CDM scale-factor, has the distinct possibility to \textit{explain} why our Universe should have expanded as it evidently has---and therein lies its greatest advantage.

\end{appendix}

\end{document}